\documentclass[times, 10pt,twocolumn]{article} 
\usepackage{latex8}
\usepackage{times}
\usepackage{graphicx}
\usepackage[none]{hyphenat}

\begin{document}

\title{Context and Keyword Extraction in Plain Text using a Graph Representation}

\author{C. Abi Chahine, N. Chaignaud, JPh Kotowicz and JP P\'ecuchet\\
INSA Rouen - LITIS EA 4108\\
Place \'Emile Blondel - BP 8\\
76131 Mont-Saint-Aignan Cedex \\
\{carlo.abi-chahine, nathalie.chaignaud, jean-philippe.kotowicz, jean-pierre.pecuchet\}@insa-rouen.fr
}

\maketitle
\thispagestyle{empty}
\begin{abstract}
Document indexation is an essential task achieved by archivists or automatic indexing tools.
To retrieve relevant documents to a query, keywords describing this document have to be carefully chosen.
Archivists have to find out the right topic of a document before starting to extract the keywords.
For an archivist indexing specialized documents, experience plays an important role.
But indexing documents on different topics is much harder.
This article proposes an innovative method for an indexing support system.
This system takes as input an ontology and a plain text document and provides as output contextualized keywords of the document.
The method has been evaluated by exploiting Wikipedia's category links as a termino-ontological resources.
\end{abstract}

\Section{Introduction}
Document indexation is an essential task achieved by archivists or automatic indexing tools.
To retrieve relevant documents to a query, keywords describing this document have to be carefully chosen. 
Archivists have to find out the right topic of a document before starting to extract the keywords. 
For a archivist indexing specialized documents, experience plays an important role. 
But indexing several documents on different topics is much harder. 
This article proposes an innovative method for an indexing support system. 
This system takes as input an ontology and a plain text document and provides as output contextualized keywords of the document.

Section  \ref{Research Context} introduces the context of our research and gives some of the state-of-the-art concerning information extraction and ontologies.
Section  \ref{Building a Graph from Words and N-grams}  and Section  \ref{Merging the Graphs}  cover our method to extract the context of the texts.
First we create graphs representing words and n-grams of the text. 
Then we merge these graphs to obtain a single consolidated weighted graph representing the text. 
Finally, in Section  \ref{Application: Wikipedia and Learning Documents}, the method is applied using Wikipedia's category links as a termino-ontological resources (TOR) and UNIT (French acronym for engineering and technology digital university) documents as a corpus of learning documents.


\Section{Research Context}
\label{Research Context}
Our aim is to design an indexing support system that assists archivists in filing records about learning documents. 
Learning documents constitute a heterogeneous corpus in terms of structure (layout of the document, format, etc.), type (lesson, exam, multiple choice questions, etc.) and topic (mathematics, physics, economics, etc.). 
The Learning Object Metadata (LOM)\footnote{Learning Object Metadata draft standard document - http://ltsc.ieee.org/wg12/20020612-Final-LOM-Draft.html,  2002. Last visited: 11/03/2008)} provides a data model describing exhaustively a learning document but it is time-consuming to fill. 
Therefore, archivists cope with lack of knowledge concerning several topics and with time constraints to fully filling the LOM records, requiring assistance to perform a complete record for a learning document.


\SubSection{Information Extraction and Document Representation}
\label{Information Extraction and Document Representation}
Information extraction (IE) consists in automatically filling databases out of texts written in natural language~\cite{Pazienza1997,Pazienza1999}.
Two activities can be discriminated:
\begin{itemize}
\item Extracting information in the text
\item Extracting information about the text
\end{itemize}
The first activity consists in extracting information and filling a database or a knowledge base with facts from the text.
Several approaches such as the dictionary-based model (e.g. Autoslog~\cite{Riloff1993}), the rule-based models (e.g. (lp)2~\cite{Ciravegna2001}), the
probabilistic models (e.g. Hidden Markov Model~\cite{Leek1997}, Conditional Random Fields~\cite{Lafferty2001}) are used to extract concepts, named entities, relations between entities and concepts, providing a reusable set of facts for other algorithms.

The second activity consists in extracting information to describe the text as a whole.
The previously mentioned methods can also be used to help in describing a text but extracting facts is not the main purpose of indexation. 
In indexation, the document is represented for special uses such as Information Retrieval (IR). Many methods and tools allow automatic indexation.


\subsubsection{Latency in a Corpus}
\label{Latency in a Corpus}
Latent methods consist in analyzing the text and extracting information without using any external knowledge.
In Natural Language Processing (NLP), Term Frequency-Inverse Document Frequency (TF-IDF)~\cite{salton1983introduction} and Latent Semantic Analysis (LSA)~\cite{Deerwester90indexingby} are commonly used to obtain significant words and documents out of a corpus.
For instance, the LSA method generates three correlation matrices out of the corpus:
\begin{itemize}
\item  WORDS x DOCUMENTS (representing the correlation between each word and each document)
\item WORDS x WORDS (representing the proximity between two words)
\item DOCUMENTS x DOCUMENTS (representing proximity between two documents)
\end{itemize}


\subsubsection{Categorization and Clustering}
\label{Categorization and Clustering}
On the one hand, Text Categorization (TC) consists in
giving to an unlabeled text a label stemming from a controlled vocabulary, using a corpus of labeled texts.
Several approaches exist for TC such as the rule-based CONTRUE system~\cite{653070} or the machine learning Naïve Bayes Classifiers (NBC)~\cite{Zheng98naivebayesian} and Support Vector Machines (SVM)~\cite{Li05svmbased}.

On the other hand, clustering breaks the corpus into clusters, ignoring any controlled vocabulary or pre-labeled data. 
Texts in the same cluster tend to be similar. 
The main difficulty linked to clustering is to find and to choose an acceptable metric space and associated distance (e.g. Euclidian distance, cosine similarity) that generate the proper clusters.

Both LSA and TC present shortcomings when it comes to retrieve from a document its contextualized keywords: the former returns uncontextualized keywords, while the latter needs training documents.


\SubSection{Ontologies and Semantic Web}
\label{Ontologies and Semantic Web}
In the early 90's, the World Wide Web (WWW) began to spread dramatically fast and the need to give web documents a semantic aspect was pressing. 
The emergence of the Semantic Web (SW) was conditioned on the development of semantic networks, which can provide meaningful relations between words (hyperonymy, hyponymy, meronymy, etc.).
For example, WordNet\footnote{WordNet - a lexical database for the english language. http://wordnet.princeton.edu/, 2006. Last visited: 11/03/2008.}~\cite{fellbaum98wordnet} is an English semantic network composed of 150 000 words, 115 000 concepts (called Synsets) and 210 000 relations.

According to~\cite{173747}, an ontology ``is an explicit specification of a conceptualization''. It is a knowledge representation where the universe (or a specific domain) is split into concepts (or classes) composed by individuals (or instances / objects) sharing relations with other individuals. 
A way of representing an ontology is the Ontology Web Language (OWL)\footnote{Web site for the speciﬁcation of OWL - http://www.w3.org/2004/OWL/, 2004. Last visited: 11/03/2008.}, a recommendation by the W3C.

There are several manners of building an ontology, but all of them are difficult to achieve. 
One can design an ontology from scratch using Knowledge Management (KM) techniques (e.g. interviewing domain experts like in Enterprise Ontology~\cite{Uschold98theenterprise}  or Methontology~\cite{Fernandez1997}) or by automatically extracting information from a corpus of texts (e.g. Terminae~\cite{SBA02}).

To provide a semantic value to our extraction and indexation, ontologies or Termino-Ontological Resources (TOR) are used such as the CISMeF Terminology\footnote{Terminologie CISMeF - http://www.chu- 
rouen.fr/terminologiecismef/. Last visited: 11/03/2008.}~\cite{DAR-00b} for medical learning documents and Wikipedia\footnote{Wikipedia, the free encyclopedia - http://www.wikipedia.org/. Last visited: 11/03/2008. }'s category links for more generic purposes.

\begin{figure}[h]
\begin{center}
\includegraphics[width=8cm]{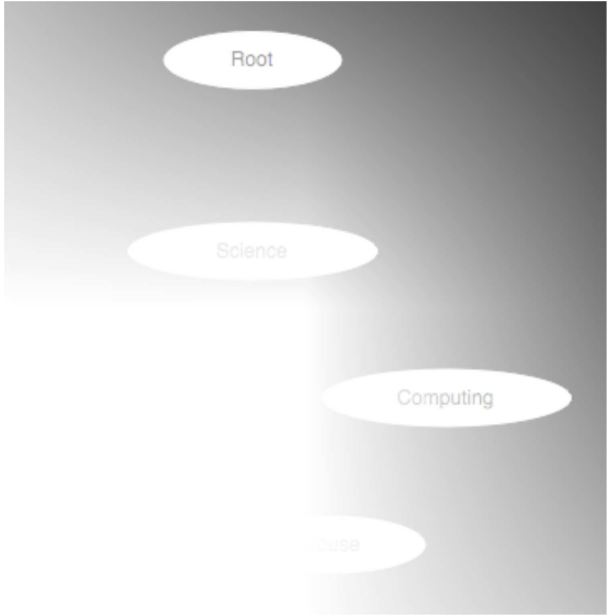}
\caption{The word ``mouse'' found in the text matches the word ``mouse'' in the terminology generating a graph.} 
\label{fig1}
\end{center}
\end{figure}

\Section{Building a Graph from Words and N-grams}
\label{Building a Graph from Words and N-grams}
Our model consists in using a TOR to create a weighted directed graph representing each word or n-gram (a sequence of n words).
The words and n-grams match the entries of the terminology (using string comparison or another metric). 
If the match is positive, the model generates a graph composed of the word or n-gram as a leaf. 
This leaf is connected by a hierarchical relation~\cite{Mekki2003} (hyponymy, meronymy, etc) to a concept node as found in the TOR. 
Concept nodes are also connected to other concept nodes by a hierarchical relation as shown in Fig. \ref{fig1}.
\Section{Merging the Graphs}
\label{Merging the Graphs}
Once the graphs of words or n-grams are built, the next step is to merge them into a consolidated graph that represents the text as a whole. 
This weighted directed graph should be configured in order to allow context and keyword extraction. 
Therefore, the weight of an edge in the graph represents the importance of the concepts within the text.

\SubSection{Affecting Weight to Edges in Graphs}
\label{Affecting Weight to Edges in Graphs}
The previous step provided unweighted directed graphs symbolizing the words/n-grams. 
A weight has to be affected to each relation (or edge) in the graphs to emphasize the importance of this relation. 
The weights affected to edges by the system have to be parameterized differently, depending on the levels they are found in. 
For example, if the context is most commonly found in the higher levels of the controlled vocabulary and/or ontology, then the weight of higher level relations is to be put forward. 
In \ref{fig1}, a larger weight is given to ``Root $\longrightarrow{}$ Science'' to promote this relation rather ``Computing $\longrightarrow{}$ Mouse''.


\SubSection{Setting a Maximum Graph Depth}
\label{Setting a Maximum Graph Depth}
A maximum depth of the graphs is needed to avoid too high level concepts or an important dissemination relatively to the original word/n-gram. 
For example, in Wikipedia there is a link between ``France'' and ``Chemistry'', which can appear far-fetched, and has a dozen levels distance.


\SubSection{The Merging Operation}
\label{The Merging Operation}
The merger of word/n-gram graphs enable the extraction of contexts or topics. 
The merging operation $\mu$ is a function that takes graph A and graph B as input and yields a new graph called $\mu(A,B)$, calculated as follows:
\begin{itemize}
\item If graph A and graph B have identical edges with respective weights $\alpha$ and $\beta$, the merged graph has the same edge with a weight $\alpha \bigoplus \beta$.
The $\bigoplus$ operation is chosen by the user of the system; it could be a simple addition operation (i.e.  $\alpha+\beta$), or a custom operation depending on various parameters (e.g. taking into account the Zipf's distribution).
\item If an edge belongs exclusively to A or B, the edge can be inserted with the same weight, the weight can be changed, or the edge can be totally ignored.
\item If A and B do not share any node, a disconnected graph can be created, graph A (or graph B) can be ignored.
\end{itemize}

The merging operation is not necessarily stationary and can vary between the steps of the computation (as applied in section V).

For instance, a text is composed of the words ``Mouse'' and ``Keyboard''.
The parameters for our example are:
\begin{itemize}
\item Weight affectation for words/n-grams: the weights of deep relations are the largest ones and are decremented level by level.
\item Maximum depth: $\infty$ (Infinity)
\item $\mu(A,B)$:
\begin{itemize}
\item $\alpha \bigoplus \beta =  \alpha + \beta$
 \item New edges are included in the merged graph with their current weights.
\item If the two graphs do not share any node, a disconnected graph is generated.
\end{itemize}
\end{itemize}

The result of the computation is shown in Fig. \ref{fig2}: 

\begin{figure}[h]
\begin{center}
 \includegraphics[width=8cm]{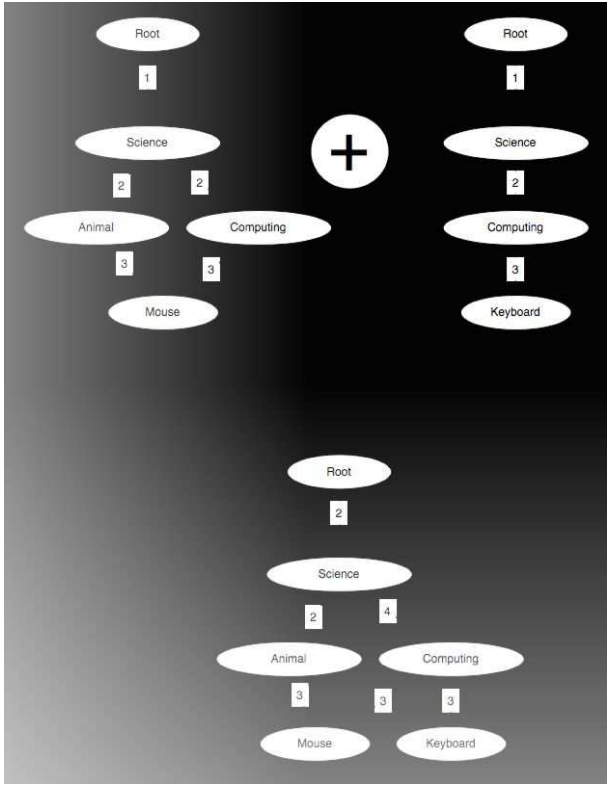}
 \caption{The parameterized merging operation and consolidated graph.} 
\label{fig2}
\end{center}
\end{figure}
As a result, the strongest relation is ``Science $\longrightarrow{}$ Computing''. 
If we only want to keep one candidate context, we can choose ``Computing'' as the main one for the text.
Therefore one way to extract contextualized keywords is to grab the words/n-grams that share ``Computing'' as an ancestor, setting up a minimal path length.
If this value is 2, ``Mouse'' and ``Keyboard'' are the contextualized keywords.


\Section{Application: Wikipedia and Learning Documents}
\label{Application: Wikipedia and Learning Documents}
Our method has been applied to Wikipedia as our TOR and controlled vocabulary and various learning documents from learning sites (UNIT\footnote{Universit\'e Num\'erique Ing\'enierie et Technologie (UNIT) - http://www.unit.eu/. Last visited: 11/03/2008.}, Wikiversity, etc.) as our corpus.

\SubSection{Wikipedia as an Ontology?}
\label{Wikipedia as an Ontology?}
Many skeptical thoughts about Wikipedia are expressed nowadays. 
As a reminder, Wikipedia is a collaborative online encyclopedia, where anyone can add an entry or fill out an existing entry. 
It is based on a Wiki (collaborative site) that contains the notion of category and sub-category. 
An article is placed by the users in one or several categories. 
As a consequence, the collaborative work gave birth to a huge TOR of categories and entries. 
To be able to use those categories as a controlled vocabulary and reliable ontology, the Wikipedia's users are considered as experts in there fields. 
In fact, for each important category, each article is carefully red by the community users, filtering out the most subjective points of view or false entries, demanding references on subjects, etc.

\begin{figure}[h]
\begin{center}
 \includegraphics[width=8.5cm,height=1.7cm]{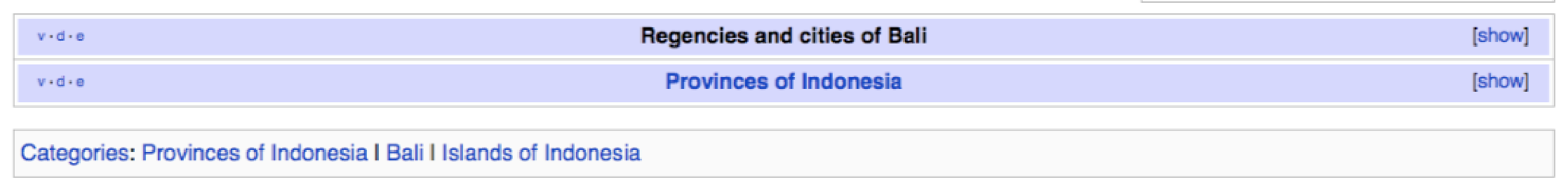}
 \caption{Wikipedia’s category link for Bali} 
\label{fig3}
\end{center}
\end{figure}

As an end user of Wikipedia, the categories of an article can be found at the bottom of the page. 
As an example, the categories of article entry ``Bali'' are ``Provinces of Indonesia'', ``Bali'' and ``Islands of Indonesia'' (Fig. \ref{fig3}). 

The category links can be downloaded under the GNU Free Documentation License. 
Another important aspect of Wikipedia database is redirection: if a concept or entry has several labels (``Dijkstra's algorithm'' and ``Dijkstra algorithm'' or ``ISO speed'' and ``film speed'') a redirection to the same article occurs.

For disambiguation purposes, Wikipedia also provides a disambiguation category avoiding homonymy and polysemy issues. 
For example Earth (classical element) is different than Earth (the default article is about planet Earth).

Moreover, Wikipedia is composed of 2 super-categories:  ``Encyclopedic Entry'' and ``Non-Encyclopedic Entry''. 
The latter has sub-categories such as ``All pages needing cleanup'' or ``Wikipedia Maintenance'' that should be ignored while indexing the documents.

Finally, links in Wikipedia are not semantically clearly defined, but they represent a hierarchical type relation (hyponymy or meronymy).


\SubSection{Learning Documents}
\label{Learning Documents}
In most of the learning objects, the educational process carries special characteristics.
A learning document, to be efficient as a learning object, tends to remind the important notions without trying to avoid repeating the important words and concepts.
Like press articles or novels for instance, the style is important and thus repetitions are avoided by using synonyms, paraphrases or pronouns. 
A second aspect in learning text is the important amount of collocations found.
The set of collocation in a technical or learning document seems to carry a greater weight in describing the context than simple words.
For example, we could find in a text ``Dikjstra algorithm'', ``Fluid mechanics'', ``Natural Language Processing'', etc.

The learning documents used in the application mostly come from UNIT and from the French version of Wikiversity. 
UNIT is an online repository of learning documents provided by French universities, institutes and engineering schools. 
The documents are manually indexed using LOM-FR (the French and European version of the LOM) by an archivist with or without the help of the documents' creators. 
Wikiversity is a wiki were anybody can add learning data about any learning topics. 
It has not as many entries as Wikipedia nor the same popularity yet. 
The learning units are indexed by categories and sub-categories (it shares the same structure as Wikipedia).

\SubSection{Parameterizing the Model}
\label{Parameterizing the Model}

Since UNIT and Wikiversity are repositories of learning documents, the n-grams (for the collocation) are fetched first as follows. 
Each time a merged graph is built, it plays the ``A'' graph role.

\begin{itemize}
\item Weight affectation for n-grams: the weights of deep relations are the largest ones and are decremented level by level.
\item Maximum depth: lower than 10 (evaluation with 7)
\item $\mu(A,B)$:
\begin{itemize}
\item $\alpha \bigoplus \beta =  \alpha + \beta$
\item New edges are included in the merged graph with their current weight,
\item If the two graphs do not share any node, a disconnected graph is generated.
\end{itemize}
\end{itemize}

The top 10 categories (10 context candidates) are picked and other nodes that do not have a path to these categories are deleted.
Then the single words are extracted.

\begin{itemize}
\item Weight affectation for single words: the weights of deep relations are the largest ones and are decremented level by level.
\item Maximum depth : lower than 5 (evaluation with 2)
\item $\mu(A,B)$ :  
\begin{itemize}
\item $\alpha \bigoplus \beta =  \alpha + \epsilon \beta$ (evaluation with $\epsilon$ = 1)
\item New edges are included in the merged graph with their current weight.
\item If the two graphs do not share any node, graph B is ignored (single word entry could create noise if they are false positives).
\end{itemize}
\end{itemize}

A model has to be chosen for ambiguous terms (homonym, polysemy).


\SubSection{Evaluating the Method and the Parameterized Model}
\label{Evaluating the Method and the Parameterized Model}
A correct evaluation process would take into account the recall, precision and f-measure scores.
Since, Wikiversity and UNIT uses a controlled vocabulary, a mapping with Wikipedia's category link has not been done yet.
This methodology is in its early stages and the only reliable way of evaluating the system is to see if the main contexts and keywords appear and if there is not too much noise.

For example, for a French lesson about software engineering, by fetching the final top 5 categories and associated keywords ordered by their individual scores, we obtained the following:
\begin{table}[ht]
 \begin{center}
   \tabcolsep = 2\tabcolsep
   \begin{tabular}{|l|l|}
   \hline
      \textbf{Contexts found} &  \textbf{Missing expected} \\
      & \textbf{contexts}\\
   \hline
Computer science & Software engineering \\
Software development &\\
Project management &\\
Information Science &\\
Formal science &\\
      \hline
      \textbf{Contextualized keywords} & \textbf{Missing expected } \\
      & \textbf{keywords}\\
   \hline
Software & \\
Software engineering & \\
Programming & \\
... & \\
... & \\
True & \\
   \hline
   \end{tabular}
\caption{Result for a French lesson about software engineering.} \label{tab1}
 \end{center}
\end{table}

Many evaluation processes share the same issues: categories tend to be too general, and keywords contain noisy results.
Consequently, we expect to have good recall for keywords but an average precision as long as free vocabulary is involved.
One way to avoid an average precision is mapping the keywords with the controlled vocabulary used by archivists.

Another observation concerns learning documents using, as a majority, examples as an educational process (e.g. C++ source codes, etc.). 
The same examples are used through out the text, giving those examples more value then the topic of the course itself.
\Section{Conclusion and Perspectives }
\label{Conclusion and Perspectives }

This method is innovative but yet needs improvements. 
In fact, with only textual information, it gives a good recall for contextualized keywords with the Wikipedia's terminology and structure. 
In order to get more precise results, it needs to be mapped to a controlled vocabulary used by archivists like the Dewey Decimal Classification (DDC) or proprietary controlled vocabulary (like the UNIT classification).

Moreover, the layouts of the documents do not play a role in the model.
Titles that are represented with a big and bold font have the same weight as ordinary text.
Yet, in learning documents, chapter titles carry a real value, in terms of context and keyword extraction.

Using Wikipedia allows us to index documents that cover a large range of topics. 
We would like to evaluate the method with a field-specific terminology and documents.
The next step will be to implement this method with the CISMeF terminology (medical terminology) and Universit\'e M\'edicale Virtuelle Francophone\footnote{Universit\'e M\'edicale Virtuelle Francophone - http://www.umvf.prd.fr/. Last visited: 11/03/2008.} (French-Speaking Virtual Medical University - UMVF) documents.

If successful, this task might enable us to accomplish another task of NLP: automatic and semiautomatic ontology mapping through a corpus of texts.
Providing contextualized keywords from two ontologies will help us correlate the concepts within a particular context thus linking together a specialist terminology (e.g. CISMeF, MeSH) with a nonspecialized terminology (e.g.Wikipedia).


\bibliography{latex8}

\begin{thebibliography}{10}\setlength{\itemsep}{-1ex}\small

\bibitem{Ciravegna2001}
F.~Ciravegna.
\newblock (lp) 2 , an adaptive algorithm for information extraction from
  web-related texts.
\newblock In {\em In Proceedings of the IJCAI-2001 Workshop on Adaptive Text
  Extraction and Mining}, 2001.

\bibitem{DAR-00b}
S.-J. Darmoni, B.~Thirion, J.~Leroy, M.~Douyere, F.~Baudic, and J.~Piot.
\newblock Cismef: a structured health resource guide for healthcare
  professionals and patients.
\newblock In {\em RIAO - Recherche d'Informations Assist\'ee par Ordinateur},
  2000.

\bibitem{Deerwester90indexingby}
S.~Deerwester, S.~T. Dumais, G.~W. Furnas, T.~K. L, and R.~Harshman.
\newblock Indexing by latent semantic analysis.
\newblock {\em Journal of the American Society for Information Science},
  41:391--407, 1990.

\bibitem{fellbaum98wordnet}
C.~Fellbaum, editor.
\newblock {\em WordNet: An Electronic Lexical Database (Language, Speech, and
  Communication)}.
\newblock The MIT Press, May 1998.

\bibitem{Fernandez1997}
M.~Fern{\'a}ndez, A.~G{\'o}mez-P{\'e}rez, and N.~Juristo.
\newblock Methontology: From ontological art towards ontological engineering.
\newblock In {\em Working Notes of the AAAI Spring Symposium on Ontological
  Engineering}, pages 33--40. Stanford University, AAAI Press, 1997.

\bibitem{173747}
T.~R. Gruber.
\newblock A translation approach to portable ontology specifications.
\newblock {\em Knowl. Acquis.}, 5(2):199--220, 1993.

\bibitem{653070}
P.~J. Hayes and S.~P. Weinstein.
\newblock Construe/tis: A system for content-based indexing of a database of
  news stories.
\newblock In {\em IAAI '90: Proceedings of the The Second Conference on
  Innovative Applications of Artificial Intelligence}, pages 49--64. AAAI
  Press, 1991.

\bibitem{Lafferty2001}
J.~Lafferty, A.~Mccallum, and F.~Pereira.
\newblock Conditional random fields: Probabilistic models for segmenting and
  labeling sequence data.
\newblock In {\em Proc. 18th International Conf. on Machine Learning}, pages
  282--289. Morgan Kaufmann, San Francisco, CA, 2001.

\bibitem{Leek1997}
T.~R. Leek.
\newblock {\em Information extraction using hidden Markov models}.
\newblock PhD thesis, UC San Diego, 1997.

\bibitem{Li05svmbased}
Y.~Li, K.~Bontcheva, and H.~Cunningham.
\newblock Svm based learning system for information extraction.
\newblock In {\em Proceedings of Sheffield Machine Learning Workshop, Lecture
  Notes in Computer Science}, pages 319--339. Springer Verlag, 2005.

\bibitem{Mekki2003}
T.~A.~E. Mekki and A.~Nazarenko.
\newblock Le r{\'e}seau terminologique, un {\'e}l{\'e}ment central pour la
  construction d'index de document.
\newblock In {\em 5i{\`e}mes rencontres Terminologie et Intelligence
  Artificielle (TIA'03)}, pages 1--10, Strasbourg, 2003.

\bibitem{Pazienza1997}
M.~T. Pazienza, editor.
\newblock {\em Information Extraction: A Multidisciplinary Approach to an
  Emerging Information Technology, International Summer School, SCIE-97,
  Frascati, Italy, 14-18, 1997}, volume 1299 of {\em Lecture Notes in Computer
  Science}. Springer, 1997.

\bibitem{Pazienza1999}
M.~T. Pazienza, editor.
\newblock {\em Information Extraction: Towards Scalable, Adaptable Systems},
  volume 1714 of {\em Lecture Notes in Computer Science}. Springer, 1999.

\bibitem{Riloff1993}
E.~Riloff.
\newblock Automatically constructing a dictionary for information extraction
  tasks.
\newblock In {\em National Conference on Artificial Intelligence}, pages
  811--816, 1993.

\bibitem{salton1983introduction}
G.~Salton.
\newblock {\em Introduction to Modern Information Retrieval (McGraw-Hill
  Computer Science Series)}.
\newblock McGraw-Hill, September 1983.

\bibitem{SBA02}
S.~Szulman, B.~Bi{\'e}bow, and N.~Aussenac-Gilles.
\newblock Structuration de terminologies {\`a} l'aide d'outils de tal avec
  terminae.
\newblock {\em Revue Traitement Automatique des Langues}, 43(1):103--128, 2002.

\bibitem{Uschold98theenterprise}
M.~Uschold, M.~King, S.~Moralee, and Y.~Zorgios.
\newblock The enterprise ontology.
\newblock {\em The Knowledge Engineering Review}, 13:31--89, 1998.

\bibitem{Zheng98naivebayesian}
Z.~Zheng.
\newblock Naive bayesian classifier committees.
\newblock In {\em Proceedings of the 10th European Conference on Machine
  Learning}, pages 196--207. Springet-Verlag, 1998.

\end{thebibliography}

\end{document}